\documentclass[12pt]{article}

\usepackage[a4paper,margin=1in]{geometry}
\usepackage{amsmath,amssymb,amsfonts}
\usepackage{bm}
\usepackage{cite}
\usepackage{hyperref}
\usepackage{setspace}
\hypersetup{
    colorlinks=true,
    linkcolor=blue,
    citecolor=blue,
    urlcolor=blue
}

\title{Electrodynamics as a Theory of Persistent Stochastic Processes}

\author{
Partha Ghose\\
\small Tagore Centre for Natural Sciences and Philosophy\\
\small Rabindra Tirtha, New Town, Kolkata 700156, India\\
\small \texttt{partha.ghose@gmail.com}
}

\date{}

\begin{document}

\maketitle

\begin{abstract}
A process-theoretic approach to electrodynamics based on persistent Kac-type
stochastic processes is developed. Finite-velocity stochastic propagation is
taken as primary, while relativistic wave equations arise as emergent
descriptions after analytic continuation of Telegrapher-type dynamics. The
Dirac and Maxwell equations are interpreted as arising from closely related
persistent propagation structures differing only in spin representation.

The framework is not intended to modify the successful empirical predictions
of quantum electrodynamics, but to provide a different underlying ontology.
Particles and fields are not treated as primitive entities with fixed intrinsic
properties. Instead, relativistic particle and field structures emerge as
stable collective modes of coupled persistent stochastic dynamics. Mass and
charge acquire interpretations respectively as persistence and stochastic
coupling scales.

Stationary bound states are interpreted as metastable persistent stochastic modes with
nontrivial internal sector dynamics. Spontaneous emission is viewed as
stochastic destabilization of such modes, while stimulated emission arises
through resonant synchronization of persistent transition currents by incident
radiation.

Gauge interactions are introduced at the level of propagation-sector
amplitudes prior to the emergence of observable probabilities. 
Radiative effects, including the anomalous magnetic moment of the electron,
are interpreted as effective stochastic dressing of coupled matter--radiation
processes. Some comments on gauge symmetry, equilibration and the Standard
Model are also included.
\end{abstract}
\section{Introduction}

Quantum mechanics and quantum electrodynamics are among the most successful
physical theories ever constructed. Their predictions for atomic and molecular
structure, condensed matter, scattering amplitudes and radiative processes have
been verified with extraordinary precision. Nevertheless, these theories remain
conceptually and mathematically incomplete, being accompanied by foundational
paradoxes as well as the occurrence of divergences requiring renormalization.
The purpose of the present paper is therefore to ask whether the equations and
effective structures of relativistic quantum theory admit a different
foundational interpretation in terms of persistent stochastic processes.

The approach developed here is inspired by two distinct lines of
thought. The first is Nelson's stochastic mechanics, in which the
nonrelativistic Schr\"odinger equation is derived from an underlying
conservative Brownian motion \cite{Nelson1966,Nelson1985}. In Nelson's
framework the wavefunction is not a fundamental postulate, but encodes the
statistical structure of a deeper stochastic process. The second is the
relativistic extension due to Gaveau, Jacobson, Kac and Schulman \cite{Gaveau1984}, who showed
that a Poisson process with finite propagation speed \cite{Kac1974}, after analytic
continuation, gives rise to the Dirac equation. In this
construction the particle propagates at speed \(c\), while random
Poisson-distributed reversals switch its propagation sector. The reversal rate
is directly related to the mass parameter after analytic continuation.

The essential conceptual shift proposed in the present paper is to take this
persistent stochastic structure seriously as a process theory. In such a
theory, the primary objects are not particles or fields with fixed intrinsic
properties, but dynamical stochastic processes with finite propagation speed,
internal sector structure, and finite correlation scales. Relativistic wave
equations then arise as effective descriptions of these processes. Physical
parameters such as mass and charge are not regarded as primitive bare
constants, but as emergent characteristics of stochastic persistence and
stochastic coupling. In this respect the programme has some affinity
with Schwinger's source theory, which sought to formulate electrodynamics in
terms of physical sources and finite response amplitudes rather than bare
operator fields \cite{SchwingerSource1,SchwingerSource2,SchwingerSource3}.

This viewpoint is especially natural for the Dirac equation and electrodynamics. Maxwell's equations
in vacuum can be written in Dirac-like form by means of the
Riemann--Silberstein vector 
\[
\mathbf F_{\pm}=\mathbf E\pm i\mathbf B.
\]
The resulting photon wave equation has the same first-order structure as a
massless relativistic wave equation \cite{Bialynicki1996}, with the spin-\(1\) matrices replacing
the spin-\(1/2\) Pauli matrices. Thus the Dirac and Maxwell equations may be
viewed as arising from a common persistent stochastic mechanism, differing only
in the spin representation carried by the process \cite{GhoseNandi2025}.

This observation suggests that the distinction between matter and radiation
may not lie in fundamentally different ontological entities, but in different
representation structures of a common persistent stochastic dynamics. The similarity is
therefore not merely mathematical, but points toward a unified
process-theoretic interpretation of relativistic quantum dynamics, a spin-\(1/2\)
persistent process giving rise to the Dirac equation \cite{Gaveau1984} and a spin-\(1\) persistent
process giving rise to the photon equation \cite{GhoseNandi2025}. The formal similarity is not
merely mathematical: it points toward a unified process-theoretic interpretation
of relativistic quantum dynamics.

A further important feature of the Kac process \cite{Kac1974} is its multi-sector probability
structure. If \(P_+\) and \(P_-\) denote the probabilities associated with the
two propagation sectors, they are not separately conserved. Probability is
continually transferred between them by the stochastic switching terms. Only
the total probability
\[
P=P_++P_-
\]
satisfies a continuity equation. This feature distinguishes persistent
stochastic mechanics from ordinary nonrelativistic Schr\"odinger mechanics,
where the probability density behaves as a single conserved fluid. It also
makes the persistent stochastic framework structurally closer to relativistic
quantum field theory, in which individual particle-number sectors are not in
general conserved, while appropriate total charges remain conserved.

The paper also revisits radiative processes from this standpoint. In standard
quantum electrodynamics, spontaneous and stimulated emission are described by
coupling atomic states to quantized radiation modes \cite{WeinbergQTF,PeskinSchroeder}. 
In the process-theoretic
picture, an excited stationary state is interpreted as a metastable persistent
stochastic mode. In the nonrelativistic diffusive limit, the persistent stochastic framework
reduces to Nelson-type stochastic mechanics. For a nondegenerate bound state,
the current velocity then vanishes while the osmotic structure remains
nontrivial, giving rise to the quantum potential. 

From the more fundamental persistent-process viewpoint, however, the stationary
state should be interpreted not merely as a diffusive stochastic equilibrium,
but as a metastable persistent propagation mode possessing internal
sector dynamics and finite correlation structure. Spontaneous emission corresponds 
to stochastic destabilization of such a
metastable persistent mode, while stimulated emission corresponds to resonant
synchronization of the persistent stochastic transition currents by an incident
radiation field.

The same viewpoint suggests a different interpretation of radiative dressing.
Instead of viewing radiative corrections as ultraviolet-divergent corrections
to fixed bare parameters, one may regard observed masses, charges and magnetic
moments as effective response parameters of coupled persistent stochastic
matter--radiation processes. 
The anomalous magnetic moment of the electron provides a particularly sharp
test case: the Schwinger term \(a_e=\alpha/2\pi\) \cite{Schwinger1948} would appear as the leading
weak-coupling approximation to a more general emergent stochastic spin
response.

This structural resemblance strengthens the possibility that persistent
stochastic process theory and quantum field theory may be observationally
equivalent descriptions of the same physical phenomena, while assigning very
different ontological meaning to the underlying structures. The point is not
that the Kac process literally reproduces particle creation and annihilation
at the outset, but that it already contains an internal sector-transfer
architecture absent from ordinary diffusion-based stochastic mechanics.

The structure of the paper is as follows. Section 2 reviews the transition
from Nelson's Brownian stochastic mechanics to Kac-type persistent stochastic
processes. It then discusses the emergence of the Dirac equation from the
Kac process and emphasizes the interpretation of mass as persistence. Section
3 reformulates Maxwell's equations using the Riemann--Silberstein vector and
shows how the photon equation fits into the same process-theoretic structure.
It also discusses the multi-sector probability structure of the Kac process
and its relation to relativistic quantum field theory. Section 4 develops the
interpretation of stationary states, osmotic motion and the quantum potential, and
applies the framework to spontaneous and stimulated emission.
Section 5 discusses radiative corrections, the Pauli term and the anomalous
magnetic moment of the electron. Section 6 comments on gauge symmetry, particle multiplets and
the Standard Model. Section 7 concludes.

\section{From Nelson Stochastic Mechanics to Persistent Processes}

\emph{\textbf{Nelson's Stochastic Mechanics}}

Nelson's stochastic mechanics begins with the hypothesis that the motion of a
microscopic particle is described not by a differentiable classical trajectory,
but by a continuous stochastic process \cite{Nelson1966,Nelson1985}. The
underlying process is taken to be a conservative Brownian motion with diffusion
coefficient
\[
\nu=\frac{\hbar}{2m}.
\]
Because Brownian paths are everywhere continuous but nowhere differentiable, one must distinguish between
forward and backward stochastic derivatives. The forward stochastic process is
written
\begin{equation}
d\mathbf x(t)
=
\mathbf b(\mathbf x,t)\,dt
+
d\mathbf W(t),
\end{equation}
where \(\mathbf b\) is the forward drift velocity and \(d\mathbf W\) is the
Wiener increment satisfying
\begin{equation}
\langle dW_i\,dW_j\rangle
=
2\nu\,\delta_{ij}\,dt.
\end{equation}
Similarly, the backward process possesses a backward drift velocity
\(\mathbf b_*\).

The current and osmotic velocities are then defined by
\begin{equation}
\mathbf v
=
\frac{1}{2}(\mathbf b+\mathbf b_*),
\qquad
\mathbf u
=
\frac{1}{2}(\mathbf b-\mathbf b_*).
\end{equation}
The probability density \(\rho(\mathbf x,t)\) satisfies the continuity equation
\begin{equation}
\partial_t\rho+\nabla\cdot(\rho\mathbf v)=0,
\end{equation}
while the osmotic velocity is related to the density by
\begin{equation}
\mathbf u
=
\nu\nabla\ln\rho.
\end{equation}
Introducing the phase function \(S(\mathbf x,t)\) through
\begin{equation}
m\mathbf v=\nabla S,
\end{equation}
one may define the complex wavefunction
\begin{equation}
\psi
=
\sqrt{\rho}\,
e^{iS/\hbar}.
\end{equation}
Nelson showed that the stochastic dynamical equations are then equivalent to
the Schr\"odinger equation
\begin{equation}
i\hbar\frac{\partial\psi}{\partial t}
=
-\frac{\hbar^2}{2m}\nabla^2\psi
+
V\psi.
\end{equation}
A central feature of Nelson's construction is that the Born probability rule is
not imposed independently. Rather, it emerges naturally from the probability
density of the underlying stochastic process. The wavefunction therefore
acquires a statistical interpretation without requiring an additional
measurement postulate.

However, the Wiener process underlying Nelson's theory possesses infinite
propagation speed. The trajectories are continuous but nowhere differentiable,
and disturbances propagate instantaneously through the stochastic medium. This
feature is acceptable in the nonrelativistic domain, but it becomes problematic
for relativistic theories where finite propagation speed is essential.

\noindent
\emph{\textbf{Persistent Kac Processes}}

A natural relativistic generalization of Brownian motion is provided by the
persistent stochastic process introduced by Kac. Unlike Wiener diffusion, the
Kac process possesses finite propagation speed. In one spatial dimension, the
particle moves with speed \(c\), but randomly reverses direction according to a
Poisson process with rate \(\lambda\).

Let \(P_+(x,t)\) and \(P_-(x,t)\) denote the probabilities for right-moving and
left-moving propagation respectively. The stochastic evolution equations are
\begin{equation}
\partial_t P_+
=
-c\,\partial_x P_+
-\lambda P_+
+\lambda P_-,
\end{equation}
\begin{equation}
\partial_t P_-
=
+c\,\partial_x P_-
+\lambda P_+
-\lambda P_-.
\end{equation}
The stochastic switching terms continuously transfer probability between the
two propagation sectors.

Defining the total probability density
\[
P=P_++P_-,
\]
and the current
\[
J=c(P_+-P_-),
\]
one obtains the continuity equation
\begin{equation}
\partial_t P+\partial_x J=0.
\end{equation}
Thus total probability remains conserved even though the individual sector
probabilities are not.

Eliminating the current yields the Telegrapher equation
\begin{equation}
\partial_t^2 P
+
2\lambda\,\partial_t P
=
c^2\partial_x^2 P.
\end{equation}
The damping term does not represent loss of total probability. Rather, it
reflects relaxation of directional persistence and continual stochastic
exchange between the propagation sectors.

The Kac process therefore differs fundamentally from ordinary diffusion. The
system possesses finite propagation speed, finite correlation time, and
intrinsic memory structure. The process is not simply diffusive, but persistent.

\noindent
\emph{\textbf{Analytic Continuation and Relativistic Wave Equations}}

A remarkable result obtained by Gaveau, Jacobson, Kac and Schulman
\cite{Gaveau1984} is that the Kac process with the addition of helicity becomes directly related to the Dirac
equation after analytic continuation of the reversal rate,
\begin{equation}
\lambda
\rightarrow
\frac{i m c^2}{\hbar}.\label{mlambda}
\end{equation}
Under this continuation, the persistent stochastic switching between the two
propagation sectors acquires oscillatory phase structure.

The resulting equations in 3-dimensions may be rewritten in spinorial form as
\begin{equation}
i\hbar\frac{\partial\psi}{\partial t}
=
m c^2
\begin{pmatrix}
0 & 1 \\
1 & 0
\end{pmatrix}
\psi
+
c
\begin{pmatrix}
\sigma_x & 0 \\
0 & -\sigma_x
\end{pmatrix}
\frac{\hbar}{i}\partial_x\psi,
\end{equation}
which is the Dirac equation in the Weyl (chiral) respresentation.

The important conceptual point is that \emph{the mass parameter arises directly from
the persistence scale of the underlying stochastic process} (\ref{mlambda}). In this picture,
mass is not introduced as a primitive intrinsic particle property. Rather, it
emerges from the stochastic reversal dynamics itself.

This observation suggests a fundamentally process-theoretic interpretation of
relativistic quantum mechanics. The relativistic wavefunction is not viewed as
a fundamental object, but as an effective description of persistent stochastic
propagation possessing finite velocity, internal sector structure and finite
correlation time.

\noindent
\emph{\textbf{Persistent Stochasticity and Relativistic Structure}}

The persistent stochastic process already possesses several structural features
normally associated with relativistic quantum field theory.

First, the propagation speed is finite from the outset. Second, the process
possesses an intrinsically multi-sector structure. Third, the individual sector
probabilities are not separately conserved. Only the total probability obeys a
continuity equation.

This differs fundamentally from ordinary Schr\"odinger mechanics, where the
probability density behaves as a single conserved fluid. Persistent stochastic
mechanics therefore possesses a richer dynamical structure even before analytic
continuation.

After analytic continuation, the sector-transfer dynamics acquires Dirac-like
form. The resulting framework shares not only the relativistic wave equations
of quantum field theory, but also part of its deeper dynamical architecture:
sector conversion, finite propagation speed and oscillatory internal dynamics.

These observations strengthen the possibility that relativistic quantum field
theory and persistent stochastic mechanics may be observationally equivalent
descriptions of the same physical phenomena, while differing fundamentally in
their ontological interpretation.

\section{Electrodynamics and the Riemann--Silberstein Representation}

\noindent
\emph{\textbf{Maxwell Equations in Dirac-like Form}}

A particularly important feature of electrodynamics is that Maxwell's equations
in vacuum may be written in a first-order form closely resembling relativistic
wave mechanics. This becomes especially transparent in the
Riemann--Silberstein representation \cite{Bialynicki1996}, in
which the electric and magnetic fields are combined into the complex vectors
\begin{equation}
\mathbf F_{\pm}
=
\mathbf E
\pm
i\mathbf B.
\end{equation}
In vacuum, Maxwell's equations are
\begin{equation}
\nabla\cdot\mathbf E=0,
\qquad
\nabla\cdot\mathbf B=0,
\end{equation}
\begin{equation}
\partial_t\mathbf E
=
c\,\nabla\times\mathbf B,
\qquad
\partial_t\mathbf B
=
-c\,\nabla\times\mathbf E.
\end{equation}
Combining these equations yields
\begin{equation}
i\partial_t\mathbf F_{\pm}
=
\pm c\,\nabla\times\mathbf F_{\pm}.
\end{equation}
The divergence-free condition becomes
\begin{equation}
\nabla\cdot\mathbf F_{\pm}=0.
\end{equation}
Introducing the spin-\(1\) matrices
\begin{equation}
(s_i)_{jk}
=
-i\epsilon_{ijk},
\end{equation}
the curl operator may be written
\begin{equation}
\nabla\times\mathbf F
=
-i\,(\mathbf s\cdot\nabla)\mathbf F.
\end{equation}
If one combines the two helicity components into a single wave function
\begin{equation}
\psi(\mathbf r,t)=
\begin{pmatrix}
F_{+}(\mathbf r,t) \\
F_{-}(\mathbf r,t)
\end{pmatrix}
\tag{50}
\end{equation}
with six components, the equation for $\psi$ in free space takes the form
\begin{equation}
i\hbar\,\partial_t \psi(\mathbf r,t)
=-i\hbar c\,({\bf \Sigma}\cdot {\mathbf \nabla })\,\psi(\mathbf r,t),
\tag{51}
\end{equation}
where
\begin{equation}
{\bf \Sigma}=
\begin{pmatrix}
{\bf s} & 0 \\
0 & -{\bf s}
\end{pmatrix}.
\tag{52}
\end{equation}
This is of the same form as the Dirac equation except for the spin structure and the absence of the mass term.

One may regard this equation also as arising from
a persistent Kac-type spin-\(1\) process. The persistent stochastic
evolution for the helicity components has the form
\begin{equation}
\partial_t F_{+} 
=
-c\,(\mathbf s\cdot\nabla)F_{+} 
-\lambda_\gamma F_{+}
+\lambda_\gamma F_{-},
\end{equation}
\begin{equation}
\partial_t F_{-}
=
+c\,(\mathbf s\cdot\nabla) F_{-}
+\lambda_\gamma F_{-}
-\lambda_\gamma F_{+} .
\end{equation}
In matrix form,
\begin{equation}
\partial_t \psi
=
-c\,\Sigma\cdot\nabla\,\psi
-\lambda_\gamma(1-\sigma_1)\psi,
\end{equation}
where
\begin{equation}
\psi
=
\begin{pmatrix}
F_{+} \\
F_{-}
\end{pmatrix},
\end{equation}
and \(\sigma_1\) acts on the helicity-sector index while the matrices \(s_i\)
act on the spin-\(1\) vector index.
The physical photon equation is recovered in the limit
\[
\lambda_\gamma\rightarrow 0,
\]
leaving only the helicity-preserving Maxwell dynamics \cite{GhoseNandi2025}. 

The distinction between matter and radiation therefore appears not as a
difference between fundamentally different physical substances, but as a
difference in the internal representation structure of the underlying
persistent stochastic process. This viewpoint suggests a common process-theoretic origin for relativistic wave
equations. Persistent stochastic propagation together with internal sector
structure generates relativistic dynamics, while the representation carried by
the process determines the observed spin-statistics character.

\noindent
\emph{\textbf{Multi-sector probability structure of the Kac process}}
The analogy becomes even more suggestive when one recalls that the Kac process
itself involves continual stochastic transfer between propagation sectors.
Although Maxwell's equations in vacuum preserve helicity in free propagation,
matter--radiation interaction may couple the helicity sectors through
stochastic interaction processes.

This suggests that the distinction between stochastic matter propagation and
electromagnetic propagation may be less fundamental than ordinarily assumed.

\noindent
\emph{\textbf{Charge and Minimal Coupling}}

In relativistic quantum theory, electromagnetic interaction is introduced
through minimal coupling. The canonical momentum is replaced according to
\[
\mathbf p
\rightarrow
\mathbf p
-
\frac{e}{c}\mathbf A.
\]
The corresponding relativistic wave equation becomes
\begin{equation}
i\hbar\partial_t\psi
=
c\,\bm{\alpha}\cdot
\left(
\hat{\mathbf p}
-
\frac{e}{c}\mathbf A
\right)\psi
+
\beta mc^2\psi
+
e\phi\,\psi.
\end{equation}
Within the present framework, this interaction is interpreted not as coupling
between fundamentally distinct fields and particles, but as stochastic coupling
between different persistent propagation processes.

The electric charge therefore acquires a different conceptual meaning. Rather
than representing a primitive intrinsic property of a particle, it becomes an
effective coupling parameter governing stochastic interaction between matter
and radiation processes.

This viewpoint suggests that charge, like mass, may ultimately arise from
deeper stochastic interaction structure rather than existing as a fundamental
bare attribute.

\section{Stationary States in Stochastic Mechanics}

In ordinary classical mechanics, a stationary bound state corresponds to the
absence of motion. In stochastic mechanics, however, the situation is
fundamentally different. Even a stationary quantum state possesses underlying
stochastic dynamics.

Consider a stationary state
\[
\psi_n(\mathbf r,t)
=
\phi_n(\mathbf r)
e^{-iE_n t/\hbar}.
\]
Writing
\[
\psi
=
e^{R+iS/\hbar},
\]
the probability density is
\[
\rho
=
|\psi|^2
=
e^{2R}.
\]
The current and osmotic velocities are
\[
\mathbf v
=
\frac{1}{m}\nabla S,
\qquad
\mathbf u
=
\frac{\hbar}{m}\nabla R.
\]
For a nondegenerate bound state, the spatial wavefunction may be chosen real.
The spatial phase is therefore constant, implying
\[
\mathbf v=0.
\]
However, unless the probability density is spatially uniform,
\[
\mathbf u\neq 0.
\]
Thus a stationary bound state is not a state of classical rest. The state
possesses nontrivial osmotic structure even though the current velocity
vanishes.

\noindent
\emph{\textbf{Quantum Potential and Osmotic Motion}}

The quantum potential is
\[
Q
=
-\frac{\hbar^2}{2m}
\frac{\nabla^2\sqrt{\rho}}{\sqrt{\rho}}.
\]
Since
\[
\sqrt{\rho}=e^R,
\]
one has
\[
\frac{\nabla^2\sqrt{\rho}}{\sqrt{\rho}}
=
\nabla^2R
+
(\nabla R)^2.
\]
Using
\[
\mathbf u
=
\frac{\hbar}{m}\nabla R,
\]
the quantum potential becomes
\[
Q
=
-\frac{m}{2}\mathbf u^2
-
\frac{\hbar}{2}\nabla\cdot\mathbf u.
\]

The quantum potential is therefore determined entirely by the osmotic
structure of the stochastic process.

For a stationary state with vanishing current velocity, the stochastic
Hamilton--Jacobi equation reduces to
\[
-E_n+V+Q=0,
\]
or
\[
V+Q=E_n.
\]
The external potential and the osmotic quantum potential therefore balance
exactly to produce the stationary bound state.

This interpretation differs sharply from the ordinary Copenhagen viewpoint.
The stationary state is not a static abstract wavefunction, but a dynamical
stochastic equilibrium maintained by nonvanishing osmotic motion.

\noindent
\emph{\textbf{Persistent Stochastic Interpretation}}

The preceding discussion refers primarily to the nonrelativistic diffusive
limit associated with Nelson stochastic mechanics. The persistent stochastic
framework permits a deeper relativistic reinterpretation of stationary states. In Nelson's Wiener process the stochastic dynamics is diffusive and Markovian.
By contrast, the Kac process possesses (i) finite propagation speed, (ii) finite correlation time,
(iii) internal propagation-sector structure, and (iv) intrinsic temporal persistence.

The relation between stationary states in Nelson stochastic mechanics and in
persistent stochastic mechanics may be clarified through the Gordon
decomposition of the conserved Dirac current
\[
j^\mu
=
\bar{\psi}\gamma^\mu\psi.
\]
The Gordon decomposition gives
\begin{equation}
j^\mu
=
\frac{i\hbar}{2m}
\left(
\bar{\psi}\partial^\mu\psi
-
(\partial^\mu\bar{\psi})\psi
\right)
+
\frac{\hbar}{2m}
\partial_\nu
(\bar{\psi}\sigma^{\mu\nu}\psi).
\end{equation}
The first term represents the convective part of the current, while the second
term represents the intrinsic spin-current contribution.

This decomposition is closely analogous to the decomposition of the stochastic
velocity in Nelson mechanics into current and osmotic parts,
\[
\mathbf v
=
\frac{1}{2}(\mathbf b+\mathbf b_*),
\qquad
\mathbf u
=
\frac{1}{2}(\mathbf b-\mathbf b_*).
\]
Within the persistent stochastic framework, the propagation-sector structure
of the Kac process naturally provides a relativistic analogue of this
decomposition. The combinations
\[
P_++P_-
\]
and
\[
P_+-P_-
\]
respectively describe total convective transport and internal sector-transfer
structure associated with stochastic switching dynamics.

The internal sector-transfer contribution therefore plays a role analogous to
the osmotic component in Nelson's stochastic mechanics. A stationary persistent
stochastic state may thus possess vanishing net convective transport while
retaining nontrivial internal sector dynamics.

From this viewpoint, relativistic stationary states are interpreted not as
static wavefunctions, but as metastable persistent stochastic structures
maintained by continual internal sector-transfer dynamics. This gives stationary 
quantum states a dynamical interpretation.
The state resembles a self-maintained stochastic resonance structure rather
than a static configuration.

\noindent
\emph{\textbf{Metastability of Excited States}}

Excited atomic states are particularly important from this viewpoint.
In standard quantum mechanics, an excited state is represented by an energy
eigenfunction. Its instability is introduced phenomenologically through
interaction with the radiation field.

In the present framework, the excited state itself is interpreted as a
metastable persistent stochastic structure. The persistence of the state is 
maintained by the internal stochastic dynamics,
while its finite lifetime reflects instability of the underlying persistent
mode.

This interpretation gives spontaneous decay a natural dynamical meaning.
Spontaneous emission corresponds to stochastic destabilization of a metastable
persistent structure rather than merely probabilistic collapse of an abstract
wavefunction.

\noindent
\emph{\textbf{Spontaneous Emission as Persistent Mode Destabilization}}

Consider an excited state
\[
\psi_e(\mathbf r,t)
=
\phi_e(\mathbf r)e^{-iE_e t/\hbar},
\]
together with a lower state
\[
\psi_g(\mathbf r,t)
=
\phi_g(\mathbf r)e^{-iE_g t/\hbar}.
\]
The transition frequency is
\[
\omega_0
=
\frac{E_e-E_g}{\hbar}.
\]
In the persistent stochastic picture, the excited state is not perfectly
stable because the underlying propagation-sector dynamics continuously
generates stochastic fluctuations in the persistent mode structure.
Spontaneous emission then corresponds to stochastic destabilization of the
excited persistent mode together with transfer of energy into the radiation
sector. Unlike ordinary classical radiation theory, the instability is intrinsic to
the persistent stochastic structure itself.

\noindent
\emph{\textbf{Stimulated Emission as Resonant Synchronization}}

Now consider the presence of incident radiation with average occupation number
\(\bar n\). The external electromagnetic field modifies the stochastic current
structure through minimal coupling,
\[
\mathbf p
\rightarrow
\mathbf p-\frac{e}{c}\mathbf A.
\]
The current velocity becomes
\[
\mathbf v
=
\frac{1}{m}
\left(
\nabla S
-
\frac{e}{c}\mathbf A
\right).
\]
Near resonance,
\[
\omega
\simeq
\omega_0,
\]
the incident field couples strongly to the persistent stochastic transition
currents associated with the excited-state decay process.

The essential point is that the incident field can synchronize the phase
structure of the stochastic transition dynamics.
From this viewpoint, stimulated emission is interpreted as resonant
synchronization of persistent stochastic transition currents.
This immediately explains the difference between spontaneous and stimulated
emission: (a) spontaneous emission is incoherent because the stochastic transition
phases are random, whereas (b) stimulated emission is coherent because the incident radiation field
phase-locks the stochastic transition process. The coherence of stimulated emission therefore does not arise from coherence of the spontaneous process itself. Rather, it arises from synchronization of the
underlying persistent stochastic dynamics by the incident radiation mode.

This interpretation is conceptually close to Bose's original insight that
probabilistic matter--radiation interaction naturally generates stimulated
radiation in thermal equilibrium \cite{Bose1924}. The present framework provides a possible
dynamical realization of that idea in terms of persistent stochastic processes.

\section{Radiative Structure and the Anomalous Magnetic Moment}

\noindent
\emph{\textbf{Radiative Corrections in Quantum Electrodynamics}}

One of the central achievements of quantum electrodynamics is the successful
description of radiative corrections. In standard perturbative QED, observable
quantities such as the electron mass, charge and magnetic moment receive
corrections arising from interaction with the quantized electromagnetic field 
\cite{WeinbergQTF,PeskinSchroeder}.

\subsection*{Gauge-Covariant Persistent Stochastic Dynamics}
While the connection between Kac-type stochastic processes \cite{Kac1974} and the Dirac
equation has been explored by Gaveau, Jacobson,
Kac and Schulman \cite{Gaveau1984}, and related stochastic path approaches have been extended to
external electromagnetic fields, a fully coupled gauge-covariant persistent
stochastic process theory in which both matter and the electromagnetic field
are represented as interacting Kac-type processes does not appear to exist in
the current literature.

The real Kac equations discussed in the preceding sections describe a genuine
persistent stochastic process for directional probabilities. However, the
introduction of gauge coupling requires a further conceptual step. Gauge
covariance acts naturally on complex amplitudes rather than directly on real
probabilities. We therefore \emph{generalize the persistent Kac structure from a
probability-level process to a sector-amplitude dynamics retaining the same
persistent switching architecture}. The probabilities are subsequently introduced as modulus squares.

Consider two particle species labelled \(a\) and \(b\). Introduce for each
species a two-component persistent sector amplitude
\begin{equation}
\Phi^{(s)}(x,t)
=
\begin{pmatrix}
\phi^{(s)}_+(x,t)\\
\phi^{(s)}_-(x,t)
\end{pmatrix},
\qquad
s=a,b.
\end{equation}
The free persistent stochastic evolution may then be written compactly as
\begin{equation}
\partial_t \Phi^{(s)}
=
-c\,\sigma_3\,\partial_x\Phi^{(s)}
-
\lambda_s
(1-\sigma_1)\Phi^{(s)},
\end{equation}
where
\begin{equation}
\sigma_1
=
\begin{pmatrix}
0&1\\
1&0
\end{pmatrix},
\qquad
\sigma_3
=
\begin{pmatrix}
1&0\\
0&-1
\end{pmatrix}.
\end{equation}
The electromagnetic interaction is introduced through the gauge-covariant
derivative
\begin{equation}
D_x^{(s)}
=
\partial_x-i e_s A_x,
\end{equation}
where \(e_s\) is the charge associated with species \(s\).
The interacting persistent stochastic equations become
\begin{equation}
\partial_t \Phi^{(s)}
=
-c\,\sigma_3 D_x^{(s)}\Phi^{(s)}
-
\lambda_s
(1-\sigma_1)\Phi^{(s)}.
\end{equation}
Explicitly,
\begin{equation}
\partial_t \phi^{(s)}_+
=
-c D_x^{(s)}\phi^{(s)}_+
-
\lambda_s\phi^{(s)}_+
+
\lambda_s\phi^{(s)}_-,
\end{equation}
\begin{equation}
\partial_t \phi^{(s)}_-
=
+c D_x^{(s)}\phi^{(s)}_-
+
\lambda_s\phi^{(s)}_+
-
\lambda_s\phi^{(s)}_-.
\end{equation}
For two species one has
\begin{equation}
D_x^{(a)}
=
\partial_x-i e_a A_x,
\qquad
D_x^{(b)}
=
\partial_x-i e_b A_x.
\end{equation}
In the physically important case of opposite charges,
\begin{equation}
e_a=e,
\qquad
e_b=-e,
\end{equation}
so that
\begin{equation}
D_x^{(a)}
=
\partial_x-i e A_x,
\qquad
D_x^{(b)}
=
\partial_x+i e A_x.
\end{equation}
The physical sector probabilities are then defined by
\begin{equation}
P^{(s)}_\pm
=
|\phi^{(s)}_\pm|^2,
\end{equation}
while the total probability density for species \(s\) is
\begin{equation}
P^{(s)}
=
|\phi^{(s)}_+|^2
+
|\phi^{(s)}_-|^2.
\end{equation}
Gauge covariance is automatic provided the amplitudes transform as
\begin{equation}
\Phi^{(s)}
\rightarrow
e^{i e_s\chi(x,t)}
\Phi^{(s)},
\end{equation}
together with
\begin{equation}
A_x
\rightarrow
A_x+\partial_x\chi.
\end{equation}
Including a scalar potential \(A_0\), one introduces the full covariant
derivatives
\begin{equation}
D_t^{(s)}
=
\partial_t+i e_s A_0,
\qquad
D_x^{(s)}
=
\partial_x-i e_s A_x.
\end{equation}
The gauge-covariant persistent stochastic equations then take the form
\begin{equation}
D_t^{(s)}\Phi^{(s)}
=
-c\,\sigma_3 D_x^{(s)}\Phi^{(s)}
-
\lambda_s
(1-\sigma_1)\Phi^{(s)}.
\end{equation}
To see explicitly how the Dirac form emerges, first rewrite the above equation as
\[
D_t^{(s)}\Phi^{(s)}
=
-c\,\sigma_3D_x^{(s)}\Phi^{(s)}
-\lambda_s\Phi^{(s)}
+\lambda_s\sigma_1\Phi^{(s)} .
\]
The scalar term \(-\lambda_s\Phi^{(s)}\) represents the overall Poisson
survival factor. It may be removed by the transformation
\[
\Phi^{(s)}(x,t)=e^{-\lambda_s t}\,\widetilde{\Phi}^{(s)}(x,t),
\]
which gives
\[
D_t^{(s)}\widetilde{\Phi}^{(s)}
=
-c\,\sigma_3D_x^{(s)}\widetilde{\Phi}^{(s)}
+\lambda_s\sigma_1\widetilde{\Phi}^{(s)} .
\]
The analytic continuation \cite{Gaveau1984}
\[
\lambda_s
\longrightarrow
-\frac{i m_s c^2}{\hbar},
\]
then yields the Dirac equation
\begin{equation}
i\hbar D_t^{(s)}\widetilde{\Phi}^{(s)}
=
-i\hbar c\,\sigma_3D_x^{(s)}\widetilde{\Phi}^{(s)}
+
m_s c^2\sigma_1\widetilde{\Phi}^{(s)}. \label{A}
\end{equation}
Thus the stochastic switching term becomes the Dirac mass term after analytic
continuation. The exponential decay factor of the real Poisson process is
thereby transformed into the oscillatory relativistic phase associated with
the rest energy.

This structure is strongly suggestive. The propagation sectors of the Kac
process behave analogously to internal spinorial degrees of freedom, while the
gauge field couples naturally through the covariant derivative prior to the
formation of probabilities. The resulting framework therefore provides a natural route toward a
gauge-covariant persistent stochastic process theory in which (i) propagation-sector amplitudes evolve stochastically,
(ii) gauge interactions enter through covariant transport,
(iii) probabilities emerge only after formation of modulus squares,
(iv) and relativistic quantum dynamics appears as an emergent collective
structure of the underlying persistent stochastic process.

\noindent
\emph{\textbf{Three-Dimensional Charged Matter Sector}}

For a physical charged spin-\(\frac12\) particle, the persistent stochastic sector
amplitude should be generalized from the one-dimensional two-sector form above to a
four-component spinor structure \cite{Gaveau1984}. Let
\begin{equation}
\Psi^{(s)}(\mathbf r,t)
=
\begin{pmatrix}
\psi^{(s)}_L(\mathbf r,t)\\
\psi^{(s)}_R(\mathbf r,t)
\end{pmatrix},
\end{equation}
where \(\psi_L\) and \(\psi_R\) are two-component Weyl spinors representing the
two chiral propagation sectors of species \(s\).
The gauge-covariant derivatives are
\begin{equation}
D_t^{(s)}
=
\partial_t+\frac{i e_s}{\hbar}A_0,
\qquad
\mathbf D^{(s)}
=
\nabla-\frac{i e_s}{\hbar c}\mathbf A .
\end{equation}
The one-dimensional Dirac equation (\ref{A}) then takes 
the form
\begin{equation}
i\hbar D_t^{(s)}\Psi^{(s)}
=
-\,i\hbar c\,\boldsymbol{\alpha}\cdot\mathbf D^{(s)}\Psi^{(s)}
+
m_s c^2\beta \Psi^{(s)} ,
\end{equation}
where
\begin{equation}
\boldsymbol{\alpha}
=
\begin{pmatrix}
-\boldsymbol{\sigma} & 0\\
0 & \boldsymbol{\sigma}
\end{pmatrix},
\qquad
\beta
=
\begin{pmatrix}
0&I_2\\
I_2&0
\end{pmatrix}.
\end{equation}
Equivalently, in terms of the two Weyl sectors,
\begin{equation}
i\hbar D_t^{(s)}\psi_L^{(s)}
=
+i\hbar c\,\boldsymbol{\sigma}\cdot\mathbf D^{(s)}
\psi_L^{(s)}
+
m_s c^2 \psi_R^{(s)},
\end{equation}
\begin{equation}
i\hbar D_t^{(s)}\psi_R^{(s)}
=
-i\hbar c\,\boldsymbol{\sigma}\cdot\mathbf D^{(s)}
\psi_R^{(s)}
+
m_s c^2 \psi_L^{(s)} .
\end{equation}
In the process-theoretic interpretation, the two Weyl components represent
persistent propagation sectors. The mass term couples these sectors and is
identified with the analytically continued Poisson switching rate,
\begin{equation}
\lambda_s
\longrightarrow
\frac{i m_s c^2}{\hbar}.
\end{equation}
Thus the inertial mass is interpreted as the persistence scale associated with
sector switching.

For two charged species \(a\) and \(b\), one writes
\begin{equation}
D_\mu^{(a)}
=
\partial_\mu+\frac{i e_a}{\hbar c}A_\mu,
\qquad
D_\mu^{(b)}
=
\partial_\mu+\frac{i e_b}{\hbar c}A_\mu,
\end{equation}
or, for opposite charges,
\begin{equation}
e_a=e,
\qquad
e_b=-e.
\end{equation}
The physical matter densities are then
\begin{equation}
\rho^{(s)}
=
\Psi^{(s)\dagger}\Psi^{(s)}
=
\psi_L^{(s)\dagger}\psi_L^{(s)}
+
\psi_R^{(s)\dagger}\psi_R^{(s)} .
\end{equation}
The conserved matter current is
\begin{equation}
j_s^\mu
=
\bar\Psi^{(s)}\gamma^\mu\Psi^{(s)} ,
\end{equation}
and the interaction with the electromagnetic process is represented at the
effective level by the usual current coupling
\begin{equation}
j_s^\mu A_\mu .
\end{equation}
Thus, in three dimensions, the charged matter sector is a gauge-covariant
persistent stochastic spin-\(\frac12\) process, while the electromagnetic
sector is a precribed gauge field. Their coupling appears,
at the emergent wave-equation level, as the standard minimal electromagnetic
coupling.

\noindent
\emph{\textbf{Including the Electromagnetic Process}}

The preceding equations treat \(A_\mu\) as an externally prescribed gauge
field. For a genuine process-theoretic formulation,  the
electromagnetic field must also be represented as a persistent stochastic
process. This can be done as in Section 3, but here we do so by writing Kac-type 
equations for a spin-\(1\) process whose sector amplitudes correspond to the helicity components of the
Riemann--Silberstein field. Analogous to Section 3, define the
six-component spin-\(1\) sector amplitude
\begin{equation}
\Psi_{\rm em}
=
\begin{pmatrix}
\mathbf F_+\\
\mathbf F_-
\end{pmatrix}.
\end{equation}
As we have seen, in the absence of sources, Maxwell's equations may be written in the
Dirac-like form \cite{Bialynicki1996}
\begin{equation}
i\hbar\,\partial_t\Psi_{\rm em}
=
c\,(\Sigma\cdot\hat{\mathbf p})\Psi_{\rm em},\label{photon}
\end{equation}
where
\begin{equation}
\Sigma_i
=
\begin{pmatrix}
s_i&0\\
0&-s_i
\end{pmatrix},
\qquad
(s_i)_{jk}=-i\epsilon_{ijk}.
\end{equation}
Schematically, introduce helicity
sector amplitudes \(\Phi^{(\gamma)}_+\) and \(\Phi^{(\gamma)}_-\), each taking
values in the spin-\(1\) representation space. The persistent stochastic
evolution has the form
\begin{equation}
\partial_t\Phi^{(\gamma)}_+
=
-c\,(\mathbf s\cdot\nabla)\Phi^{(\gamma)}_+
-\lambda_\gamma \Phi^{(\gamma)}_+
+\lambda_\gamma \Phi^{(\gamma)}_-,
\end{equation}
\begin{equation}
\partial_t\Phi^{(\gamma)}_-
=
+c\,(\mathbf s\cdot\nabla)\Phi^{(\gamma)}_-
+\lambda_\gamma \Phi^{(\gamma)}_+
-\lambda_\gamma \Phi^{(\gamma)}_- .
\end{equation}
In matrix form,
\begin{equation}
\partial_t\Phi^{(\gamma)}
=
-c\,\Sigma\cdot\nabla\,\Phi^{(\gamma)}
-\lambda_\gamma(1-\sigma_1)\Phi^{(\gamma)},
\end{equation}
where
\begin{equation}
\Phi^{(\gamma)}
=
\begin{pmatrix}
\Phi^{(\gamma)}_+\\
\Phi^{(\gamma)}_-
\end{pmatrix},
\end{equation}
and \(\sigma_1\) acts on the helicity-sector index while the matrices \(s_i\)
act on the spin-\(1\) vector index.
The photon equation (\ref{photon}) is recovered in the limit
\begin{equation}
\lambda_\gamma\rightarrow 0,
\end{equation}
leaving only the helicity-preserving Maxwell dynamics \cite{GhoseNandi2025}. In this sense the
electromagnetic field is not merely an externally imposed gauge potential, but
a spin-\(1\) persistent stochastic process whose emergent wave equation is
Maxwell's equation.

Matter--radiation interaction is then described by coupling the spin-\(1/2\)
matter process to this spin-\(1\) electromagnetic process. At the effective
level this coupling appears as the usual gauge-covariant replacement
\begin{equation}
\partial_\mu\rightarrow D_\mu=\partial_\mu-i e A_\mu,
\end{equation}
but in the process-theoretic interpretation \(A_\mu\) is itself an emergent
collective variable associated with the spin-\(1\) persistent stochastic
process.

The electromagnetic interaction is introduced at the level of persistent
propagation-sector amplitudes rather than directly at the level of observable
probabilities. The gauge-covariant stochastic dynamics therefore modifies the
sector amplitudes themselves, and the physical probabilities emerge only after
formation of modulus squares. \emph{At the emergent level, the resulting collective
dynamics reproduces the standard Dirac and Maxwell equations together with
their familiar electromagnetic couplings}.

From this viewpoint, quantities such as the electron mass and charge are not
fundamental microscopic parameters inserted into the theory from the outset.
Rather, they are effective collective characteristics of the coupled
matter--radiation persistent stochastic process. \emph{The observed relativistic
particle and field structures therefore emerge from the stabilization of the
underlying stochastic sector dynamics}.

\noindent
\emph{\textbf{Anomalous Magnetic Moment as Stochastic Spin Dressing}}

Within the persistent stochastic framework, the free spin-\(\frac12\) Kac
process naturally reproduces the Dirac magnetic moment with gyromagnetic ratio
\[
g=2.
\]
This value reflects the intrinsic spinorial transport structure of the free
persistent process itself.

The anomalous magnetic moment arises only after coupling the spin-\(\frac12\)
matter process to the spin-\(1\) electromagnetic persistent process. The
interaction modifies the internal propagation-sector dynamics and the
associated spin-current structure appearing in the Gordon decomposition of the
Dirac current. The resulting collective stochastic dressing produces a small
correction to the effective magnetic response.

From this viewpoint, the anomaly
\[
a_e=\frac{g-2}{2}
\]
does not represent the self-energy correction of a bare point particle, as in quantum 
electrodynamics. Rather, it measures the finite modification of the effective spin transport
generated by repeated stochastic matter--radiation interaction.

The Schwinger value 
\[
a_e=\frac{\alpha}{2\pi}
\]
is then interpreted as the leading weak-coupling approximation to this
collective stochastic spin dressing. The full observed anomalous magnetic
moment reflects the complete structure of the coupled persistent stochastic
matter--radiation process.

At the present stage, it would be premature to claim that the stochastic
framework derives the observed anomaly without any analogue of parameter
renormalization. What can be said is that the anomaly may be interpreted not
as a shift of the defining process parameters \(m\) and \(e\), but as an
emergent finite response coefficient generated by the coupled stochastic
matter--radiation dynamics.

This viewpoint changes the conceptual meaning of radiative corrections.
Instead of:
\[
\text{bare particle}
+
\text{divergent vacuum self-energy},
\]
one has:
\[
\text{persistent stochastic process}
\rightarrow
\text{effective finite dynamical response}.
\]
The observable magnetic moment is therefore interpreted not as the property of
an isolated particle modified by vacuum fluctuations, but as an emergent
effective characteristic of coupled stochastic propagation and interaction.

\noindent
\emph{\textbf{Relation to Schwinger's Source Theory}}

The present programme bears some conceptual similarity to Schwinger's source
theory \cite{SchwingerSource1,SchwingerSource2,SchwingerSource3}.
Schwinger sought to formulate electrodynamics directly in terms of physically
observable sources and finite transition amplitudes, avoiding the ontological
primacy of operator-valued quantum fields and the associated emphasis on bare
particles and vacuum divergences.

Similarly, the present framework places primary emphasis not on quantized
operator fields, but on persistent stochastic propagation processes and their
effective observable responses.

The analogy should not be overstated. Persistent stochastic mechanics is not a
reformulation of source theory. Nevertheless, both approaches suggest that the
successful empirical structure of quantum electrodynamics may admit a deeper
interpretation in terms of finite physical processes rather than divergent bare
entities.
.

\section{Gauge Symmetry, Particle Multiplets and the Standard Model}

\noindent
\emph{\textbf{Symmetry in Quantum Theory}}

Symmetry principles play a central role in modern quantum theory. The Standard
Model is constructed around local gauge symmetries together with internal group
representations that organize the observed particle multiplets \cite{WeinbergQTF,PeskinSchroeder}.

Within conventional quantum field theory, these symmetries are introduced at
the level of operator-valued fields and local gauge invariance. Renormalizable
interactions are then constructed by requiring invariance under the relevant
symmetry groups. The present process-theoretic framework does not deny the empirical success of
this structure. Rather, it raises the possibility that the observed symmetry
structure may itself emerge from deeper properties of persistent stochastic
propagation.

\noindent
\emph{\textbf{Symmetry as Structure of Persistent Processes}}

In the persistent stochastic picture, the fundamental entities are not
particles or quantum fields, but propagation processes possessing internal
sector structure. The emergence of relativistic wave equations from persistent stochastic
dynamics already suggests that representation theory enters naturally through
the internal structure of the process itself. Different particle types may therefore 
correspond to different persistent
stochastic representations: spin-\(1/2\) representations generate Dirac-type processes,
spin-\(1\) representations generate Maxwell-type processes, and
more complicated internal symmetry structures may generate multiplet
structure analogous to that appearing in the Standard Model.

From this viewpoint, particle multiplets need not be regarded as fundamentally
distinct collections of particles. Instead, they may correspond to different
collective modes or representations of underlying persistent stochastic
dynamics.

\noindent
\emph{\textbf{Gauge Symmetry and Stochastic Coupling}}

Gauge symmetry occupies a special place in conventional quantum field theory.
In standard formulations, local gauge invariance determines the interaction
structure and plays an essential role in renormalizability.
In the present framework, however, renormalizability is not the primary
guiding principle. The emphasis shifts toward finite stochastic interaction
structure and effective dynamical response.

This raises the possibility that gauge symmetry itself may acquire a different
interpretation. Rather than being a fundamental principle imposed at the level of operator
fields, gauge structure may emerge as an effective large-scale invariance of
the underlying persistent stochastic dynamics.

The appearance of minimal coupling in electrodynamics, an Abelian gauge
theory, is interpreted in the present framework not merely as a formal
replacement rule, but as the effective large-scale manifestation of stochastic
coupling between persistent matter and radiation processes. In this sense, the
Abelian gauge interaction is already incorporated process-theoretically through
coupled propagation-sector amplitude dynamics. This naturally raises the
question of whether non-Abelian gauge interactions may likewise emerge from
more general internal stochastic sector structures and collective coupling
symmetries of persistent processes.
Gauge invariance would then reflect redundancy in the effective large-scale
description of these coupled stochastic dynamics rather than a fundamental
property of bare microscopic fields. 

At present this idea remains speculative. Nevertheless, the process-theoretic
framework naturally invites such reinterpretation because the fundamental
ontology differs so radically from that of ordinary local quantum field theory.

\noindent
\emph{\textbf{Spontaneous Symmetry Breaking and the Higgs Boson}}

A similar reinterpretation may apply to spontaneous symmetry breaking.
In the Standard Model, the Higgs field acquires a nonzero vacuum expectation
value, generating particle masses through spontaneous breaking of electroweak
symmetry. Within the present framework, however, mass already acquires a direct
dynamical interpretation through persistent stochastic sector-switching, and 
renormalization is not an issue.

This suggests that the Higgs mechanism itself may represent an effective
collective description of deeper stochastic persistence structure.
The Higgs boson would then be interpreted not necessarily as a fundamental
elementary scalar field, but as a collective excitation associated with
symmetry-breaking properties of an underlying stochastic medium.

Such a viewpoint would resemble the appearance of collective excitations in
condensed matter systems, where effective quasiparticles emerge from deeper
many-body dynamics.

The present paper does not attempt to construct such a theory explicitly.
Rather, the point is conceptual: once relativistic quantum dynamics is
reinterpreted process-theoretically, the Standard Model symmetry structure may
itself admit a deeper dynamical interpretation.

\noindent
\emph{\textbf{Observational Equivalence and Ontology}}

The central claim of the present programme is not that the Standard Model or
quantum electrodynamics are empirically incorrect. On the contrary, the remarkable 
success of relativistic quantum field theory strongly suggests that any viable 
process-theoretic description must reproduce its effective large-scale predictions 
to extremely high precision.

The proposal is instead that quantum field theory may represent an effective
emergent description of deeper persistent stochastic propagation processes.
From this viewpoint
(i) relativistic wave equations emerge from persistent stochastic dynamics,
(ii) mass reflects persistence scale, (iii) charge reflects stochastic coupling strength,
(iv) radiative structure reflects effective stochastic dressing, and (v) symmetry 
structure reflects collective properties of persistent
stochastic propagation. The observational content of relativistic quantum field theory may therefore be
retained while the underlying ontology is radically transformed.

This possibility is especially intriguing because the persistent stochastic
framework already reproduces several structural features ordinarily associated
with relativistic quantum field theory: (i) finite relativistic propagation speed, (ii)  
multi-sector dynamics, (iii) sector-transfer structure, (iv) Dirac-type relativistic evolution,
and (v) coupled matter--radiation interaction.

These similarities strengthen the possibility that persistent stochastic
mechanics and relativistic quantum field theory may be observationally
equivalent descriptions of the same physical phenomena while differing
fundamentally in their interpretation of physical reality.

\section{Conclusion}
The framework proposed reproduces several deep structural
features of relativistic quantum theory while providing a radically different
physical interpretation. Relativistic wave equations emerge from persistent
stochastic propagation; the Dirac and Maxwell equations arise from closely
related spin-\(\frac12\) and spin-\(1\) persistent processes; and gauge
interactions are introduced at the level of propagation-sector amplitudes prior
to the emergence of observable probabilities.

Within this viewpoint, the parameters \(m\) and \(e\) appearing in the
stochastic equations play the role of defining process parameters associated
respectively with persistence scale and stochastic coupling strength. The
observable relativistic particle and field structures then emerge as stable
collective modes of the coupled stochastic dynamics. Radiative structure,
including the anomalous magnetic moment, is interpreted not as divergent
renormalization of bare point-particle entities, but as effective stochastic
dressing of the internal propagation-sector dynamics generated by repeated
matter--radiation interaction.

The present work therefore suggests the possibility that relativistic quantum
field theory may ultimately represent an emergent effective description of
deeper persistent stochastic process dynamics.

\section{Acknowledgement}

I acknowledge use of ChatGPT for language polishing.


\begin{thebibliography}{99}

\bibitem{Nelson1966}
E.~Nelson,
``Derivation of the Schr\"odinger Equation from Newtonian Mechanics,''
\emph{Physical Review} \textbf{150}, 1079--1085 (1966).

\bibitem{Nelson1985}
E.~Nelson,
\emph{Quantum Fluctuations},
Princeton University Press, Princeton (1985).

\bibitem{Gaveau1984}
B.~Gaveau, T.~Jacobson, M.~Kac and L.~S.~Schulman,
``Relativistic Extension of the Analogy Between Quantum Mechanics and Brownian Motion,''
\emph{Physical Review Letters}
\textbf{53}, 419--422 (1984).

\bibitem{Kac1974}
M.~Kac,
``A Stochastic Model Related to the Telegrapher's Equation,''
\emph{The Rocky Mountain Journal of Mathematics}
\textbf{4}, 497--509 (1974).


\bibitem{SchwingerSource1}
J.~Schwinger,
\emph{Particles, Sources, and Fields, Vol.~I},
Addison--Wesley, Reading, Massachusetts (1970).

\bibitem{SchwingerSource2}
J.~Schwinger,
\emph{Particles, Sources, and Fields, Vol.~II},
Addison--Wesley, Reading, Massachusetts (1970).

\bibitem{SchwingerSource3}
J.~Schwinger,
\emph{Particles, Sources, and Fields, Vol.~III},
Addison-Wesley (1989).

\bibitem{Bialynicki1996}
I.~Bia{\l}ynicki-Birula,
``Photon Wave Function,''
in \emph{Progress in Optics}, Vol.~36,
edited by E.~Wolf,
Elsevier, Amsterdam (1996), pp.~245--294

\bibitem{GhoseNandi2025}
P. Nandi and P. Ghose,
``Stochastic Quantization of Electrodynamics and Linearized Gravity,''
arXiv:2508.10190 [gr-qc] (2025).

\bibitem{WeinbergQTF}
S.~Weinberg,
\emph{The Quantum Theory of Fields, Vols.~I \& 2},
Cambridge University Press, Cambridge (1995).

\bibitem{PeskinSchroeder}
M. E. Peskin and D. V. Schroeder,
\emph{An Introduction to Quantum Field Theory}, Addison-Wesley (1993). 

\bibitem{Schwinger1948}
J.~Schwinger,
``On Quantum-Electrodynamics and the Magnetic Moment of the Electron,''
\emph{Physical Review}
\textbf{73}, 416--417 (1948).

\bibitem{weyl}
H. Weyl,`Elektron und Gravitation.I', {\em Z. Phys.} {\bf 56}, 330-352 (1929).

\bibitem{Bose1924}
S.~N.~Bose,
``W\"{a}rmegleichgewicht 
im Strahlungsfeld bei Anwesenheit von Materie,'' Z. Phys. \textbf{27}, 384--393 (1924).

\end{thebibliography}
\end{document}